\documentclass[aps,prd,nofootinbib,twocolumn,superscriptaddress,floatfix,notitlepage]{revtex4-1}
\pdfoutput=1

\usepackage{graphicx}
\usepackage{amsmath,amsfonts,amssymb}
\usepackage{color}
\usepackage[breaklinks,colorlinks,urlcolor=blue,citecolor=blue,linkcolor=magenta]{hyperref}
\usepackage{verbatim}
\usepackage{enumitem}
\usepackage{hyperref}

\newcommand{\td}{{\rm d}}
\newcommand{\vect}[1]{\boldsymbol{#1}}

\newcommand{\be}{\begin{equation}}
\newcommand{\ee}{\end{equation}}
\newcommand{\bea}{\begin{equation} \begin{aligned}}
\newcommand{\eea}{\end{aligned} \end{equation}}

\def\lsim{\mathrel{\raise.3ex\hbox{$<$\kern-.75em\lower1ex\hbox{$\sim$}}}}
\def\gsim{\mathrel{\raise.3ex\hbox{$>$\kern-.75em\lower1ex\hbox{$\sim$}}}}

\begin{document}

\title{Gravitational wave microlensing by dressed primordial black holes}

\author{Juan Urrutia}
\email{juan.urrutia@kbfi.ee}
\affiliation{Keemilise ja Bioloogilise F\"u\"usika Instituut, R\"avala pst. 10, 10143 Tallinn, Estonia}
\affiliation{Departament of Cybernetics, Tallinn University of Technology, Akadeemia tee 21, 12618 Tallinn, Estonia}
\author{Ville Vaskonen}
\email{ville.vaskonen@pd.infn.it}
\affiliation{Keemilise ja Bioloogilise F\"u\"usika Instituut, R\"avala pst. 10, 10143 Tallinn, Estonia}
\affiliation{Dipartimento di Fisica e Astronomia, Universit\`a degli Studi di Padova, Via Marzolo 8, 35131 Padova, Italy}
\affiliation{Istituto Nazionale di Fisica Nucleare, Sezione di Padova, Via Marzolo 8, 35131 Padova, Italy}
\author{Hardi Veerm\"ae}
\email{hardi.veermae@cern.ch}
\affiliation{Keemilise ja Bioloogilise F\"u\"usika Instituut, R\"avala pst. 10, 10143 Tallinn, Estonia}

\begin{abstract}
   We study gravitational wave microlensing by primordial black holes (PBHs), accounting for the effect of a particle dark matter minihalo surrounding them. Such minihaloes are expected when PBHs make up only a fraction of all dark matter. 
  We find that current LIGO-Virgo detections disfavor dark matter in the form of PBHs heavier than $100 M_{\odot}$ at $1.4\sigma$.
   The next generation observatories can potentially probe PBHs as light as $0.01 M_\odot$ and down to $2\times10^{-4}$ fraction of all dark matter. We also show that these detectors can distinguish between dressed and naked PBHs, providing a novel way to study the distribution of particle dark matter around black holes and potentially shed light on the origins of black holes.
\end{abstract}

\maketitle

\section{Introduction}

Primordial black holes (PBHs) as a potential dark matter (DM) candidate, have gained renewed interest due to their testability through gravitational wave (GW) observations~\cite{Nakamura:1997sm, Eroshenko:2016hmn}. Given the existing constraints on PBH~\cite{Carr:2020gox}, they may comprise all of DM only in the asteroid mass window $10^{-16} M_\odot \lesssim m_{\rm PBH} \lesssim 10^{-11} M_\odot$. Yet, heavier PBHs may be related to the seeding of cosmic structures~\cite{1983ApJ...275..405F,1983ApJ...268....1C, Carr:2018rid} including the high redshift surprisingly luminous galaxies observed by the James Webb telescope~\cite{Liu:2022bvr, Hutsi:2022fzw, Yuan:2023bvh}.

After the first detections of black hole (BH) binaries by LIGO~\cite{LIGOScientific:2016aoc}, speculations of their possible primordial origin were presented~\cite{Sasaki:2016jop, Bird:2016dcv, Clesse:2016vqa}. The subsequent analyses of the observed binary population~\cite{LIGOScientific:2018mvr, LIGOScientific:2020ibl, LIGOScientific:2021djp} indicate that many of these BHs are likely to have an astrophysical origin~\cite{Hutsi:2020sol, Hall:2020daa, Franciolini:2021tla, He:2023yvl}, while the observed merger rate suggests that stellar mass PBHs cannot account for more than a percent of all DM~\cite{Raidal:2017mfl, Raidal:2018bbj, Vaskonen:2019jpv, Hutsi:2020sol, Wong:2020yig, DeLuca:2020jug, Franciolini:2022tfm}. The next-generation GW observatories can probe PBH binary populations across a broad parameter range~\cite{Pujolas:2021yaw, Barsanti:2021ydd}.

Gravitational lensing has provided important probes of PBH DM, with $10^{-11} M_\odot \lesssim m_{\rm PBH} \lesssim 30 M_\odot$ PBHs constrained by stellar microlensing~\cite{Macho:2000nvd, EROS-2:2006ryy, Griest:2013aaa, Niikura:2017zjd, Smyth:2019whb, Niikura:2019kqi, Oguri:2022fir, Cai:2022kbp} and heavier PBHs by the lensing of type Ia supernovae~\cite{Zumalacarregui:2017qqd, Dhawan:2023ekc} or GWs~\cite{Urrutia:2021qak}. At high masses, $m_{\rm PBH} \gtrsim 100 M_\odot$, the most stringent constraints arise from the accretion of baryons into PBHs~\cite{Ricotti:2007au, Horowitz:2016lib, Ali-Haimoud:2016mbv, Poulin:2017bwe, Hektor:2018qqw, Hutsi:2019hlw, Serpico:2020ehh}. 

The search for PBHs and other compact astrophysical objects in the stellar mass range can be conducted through GW lensing, with next-generation GW observatories, such as the Einstein Telescope (ET), having the potential to confirm or exclude the primordial origin for the observed BH mergers~\cite{Jung:2017flg, Diego:2019rzc, Liao:2020hnx, Urrutia:2021qak, Basak:2021ten, Zhou:2022yeo}. Conventional optical microlensing searches rely on the lens transiting through its Einstein radius and become challenging when the transit time exceeds the duration of the experiment. GW microlensing, on the other hand, relies on the interference of the multiple paths the GW takes around the lens~\cite{Takahashi:2003ix} allowing for the detection of much heavier lenses. Although various different DM substructures have been considered as lenses~\cite{Diego:2019lcd, Oguri:2020ldf, Choi:2021bkx, Guo:2022dre, Tambalo:2022wlm}, the expected rate of such events has been shown to be low~\cite{Fairbairn:2022xln}.

In this paper, we constrain the abundance of PBHs using GW microlensing and the LIGO-Virgo observations and examine the capabilities of current and future GW detectors, including LIGO O5 (A+)~\cite{LIGOAplus} and ET~\cite{Hild:2010id}. We include the effect of a particle dark matter (PDM) minihalo surrounding the PBHs, which is expected to form when PBHs make up only a fraction of DM. These PDM minihaloes have been extensively studied as they are relevant for precise predictions of lensing surveys, WIMP annihilation signals, and PBH accretion~\cite{Bertschinger:1985pd, Mack:2006gz, Ricotti:2007au, Eroshenko:2016yve, Adamek:2019gns, Inman:2019wvr, Carr:2020mqm, Serpico:2020ehh, DeLuca:2020bjf, Boudaud:2021irr, Gines:2022qzy, Oguri:2022fir, Cai:2022kbp}.

Our work incorporates several key improvements when compared to existing literature. On top of considering the well-motivated effect of the DM dress around the PBHs in the context of GW lensing, we do so in full detail without relying on approximations of the lens profile. We use GW waveform templates that include the inspiral, merger and ringdown phases, and employ a log-likelihood analysis to accurately assess the detectability of lensing. These considerations allow us to place reliable constraints on the PBH abundance from existing LIGO-Virgo data and to derive prospects for the upcoming LIGO observational runs and ET. Additionally, we discuss the ability of the next-generation GW detectors to differentiate between BH environments. We show that ET will be able to identify the presence of DM minihalos around BHs. This would provide a new way to gain insight into the origins of binary black hole populations that is independent of the merger rate and could also shed light on the nature of DM.

\section{DM dress of PBHs}

\begin{figure}
\centering
\includegraphics[width=0.85\columnwidth]{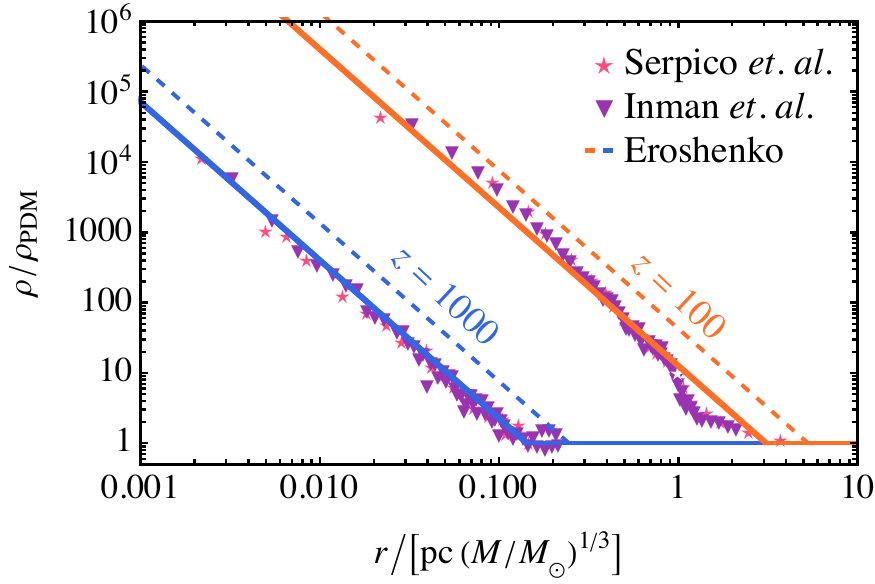}
\caption{The PDM minihalo profile~\eqref{eq:profile} (solid lines) at redshifts $z=1000$ and $z=100$ compared to the estimate in Ref.~\cite{Eroshenko:2016yve} (dashed lines) and profiles of isolated PBHs obtained via $N$-body simulations~\cite{Inman:2019wvr,Serpico:2020ehh} (points).}
\label{fig:profile}
\end{figure}

For GW microlensing, the details of the inner profile slope are not crucial, and we can adapt the density profile~\cite{Adamek:2019gns,Boudaud:2021irr}\footnote{Geometric units $c=G=1$ are used throughout this paper.}
\bea \label{eq:profile}
    &\rho_h(r) = \rho_{\rm PDM, eq} \left(\frac{r}{r_{\rm eq}}\right)^{-\!\frac94} \theta(R_h-r) \\
    &\approx \frac{0.4 M_\odot}{{\rm pc}^3} (1-f_{\rm PBH}) \left[ \frac{m_{\rm PBH}}{M_\odot} \right]^{\!\frac34} \left[ \frac{r}{\rm pc} \right]^{-\!\frac94} \theta(R_h-r) ,
\eea
where $\rho_{\rm PDM, eq} = (1-f_{\rm PBH}) \rho_{\rm eq} \Omega_{\rm DM}/\Omega_{\rm M}$ is the average proper cold particle DM energy density at matter-radiation equality, $f_{\rm PBH}$ is the fraction of DM in the form of PBHs, $r_{\rm eq} = 0.4(m_{\rm PBH} \rho_{\rm eq})^{1/3}$ is the turnaround radius at matter-radiation equality, and $R_h$ denotes the proper size of the minihalo. As can be seen in Fig.~\ref{fig:profile}, this profile is in good agreement with the results of $N$-body simulations of isolated PBHs~\cite{Inman:2019wvr,Serpico:2020ehh}, but it predicts a less dense and thus a less massive PDM minihalo than was estimated by approximate analytic arguments~\cite{Mack:2006gz,Ricotti:2007au,Eroshenko:2016yve}.

To determine $R_h$, we first note that the PDM minihalo mass $M_h$ is limited by the available amount of PDM in the universe. This implies that $f_{\rm PBH} M_{h} \leq (1 - f_{\rm PBH}) m_{\rm PBH}$. This bound is saturated in the idealized case when all of the PDM is contained in PBH minihaloes. A stronger condition for $R_h$, however, arises from the formation of DM structures --  as the minihaloes grow via secondary accretion in the matter-dominated phase~\cite{Bertschinger:1985pd}, one expects that the minihalo growth is halted once the PBHs are absorbed into non-linear DM structures. \footnote{ Virial velocities can be quite large even in small early PBHs clusters and stifle the binding of PDM to PBHs. In general, minihalo formation is poorly understood in this regime. Thus, to be conservative, we make the assumption that the minihalo growth stops at the onset of non-linear structure formation.} If $f_{\rm PBH}$ is large, the PBHs themselves can begin to form small clusters of a few PBHs at the onset of matter domination~\cite{Raidal:2018bbj, Hutsi:2019hlw, Inman:2019wvr, DeLuca:2020jug}. We estimate that this effect stops the minihalo growth when half of the PBHs have formed structures containing two or more PBHs. This happens at $1+z_{\rm NL} \approx 5500 f_{\rm PBH}$~\cite{Inman:2019wvr}. Additionally, we expect that the usual CDM cosmology holds at larger scales so that non-linear structure formation becomes relevant not later than at redshift 30, which sets the lower bound on $z_{\rm NL}$. Defining the size of the PDM minihalo by the radius at which the density of the minihalo matches the ambient PDM density, we find that $\rho_{h}(R_{h}) \approx \rho_{\rm PDM}(z_{\rm NL})$, which gives
\be\label{radius_halo}
    R_{h} \approx 14\,{\rm pc} \, \left[ \frac{m_{\rm PBH}}{M_\odot} \right]^{\!\frac13} \, \min\!\left(1,10^{-3} f_{\rm PBH}^{-\frac43} \right) \,.
\ee
Consequently, the minihalo mass is
\be~\label{eq:Mh}
    M_h \approx (1-f_{\rm PBH}) m_{\rm PBH} \, \min\!\left(50, 0.28 f_{\rm PBH}^{-1} \right)\,.
\ee
Therefore, the PDM minihalo can be at most 50 times heavier than the central BH, and at most 28\% of PDM can be bound to PBHs.

The minihalo profile is not well understood for large $f_{\rm PBH}$ and at small $z$. $N$-body simulations~\cite{Inman:2019wvr} indicate that the minihalo around isolated PBHs becomes steeper at larger $f_{\rm PBH}$, potentially due to the tidal effects that the neighboring PBHs have on the outer shells of the PDM minihalo. On the other hand, clusters of several PBHs possess much shallower PDM minihaloes. In this paper, we consider the region of the density profile that was formed when most PBHs were isolated. By the above arguments, most PBHs were isolated at $z \gtrsim 100$ when $f_{\rm PBH} \gtrsim 0.02$, that is, at the epoch when the profile~\eqref{eq:profile} has been numerically tested. Moreover, if $f_{\rm PBH} \gtrsim 0.2$, the PBHs are heavier than the minihalos surrounding them, so the GW microlensing is only mildly affected by the latter. In summary, we expect that a more detailed description of the PBH dressing would not strongly affect our results.

\section{Microlensing formalism}

We consider a compact object binary emitting GWs at the angular diameter distance $D_s$, which, on their way to the detector, pass by a PBH surrounded by a DM minihalo at the angular diameter distance $D_l$. The microlensing effect on the Fourier transform of the GW signal $\tilde \phi(f)$ can be characterized by the amplification factor $F(f)$ so that the Fourier transform of the lensed GW signal is $\tilde\phi_L(f) = F(f) \tilde\phi(f)$. In the thin-lens and point source \footnote{As shown in~\cite{Matsunaga:2006uc} the point source approximation is very good in the parameter range considered in this work.} approximation~\cite{schneider2012gravitational},
\be \label{eq:F}
    F(w,\vect{y}) = \frac{w}{2i\pi} \int \td^2 x \,e^{i w T(\vect{x},\vect{y})} \,,
\ee
where the integral is over the lens plane and $\vect{y}$ is a vector in the lens plane that determines the displacement of the source from the line of sight to the lens. The dimensionless frequency $w$ and the dimensionless time delay function $T$ are
\bea
    &w = 2\pi f \frac{(1+z_l) D_s}{D_l D_{ls}} \xi_0^2  \,, \\
    &T(\vect{x},\vect{y}) = \frac12 |\vect{x}-\vect{y}|^2 - \psi(\vect{x}) - \phi_m(\vect{y}) \,,
\eea
where $D_{ls} = D_s - D_l (1+z_l)/(1+z_s)$, $\xi_0$ is a characteristic lens scale, $\psi(\vect{x})$ denotes the deflection potential and the function $\phi_m(\vect{y})$ is defined so that the minimum of $T(\vect{x},\vect{y})$ for a fixed $\vect{y}$ is zero. The lens potential is determined by the density profile of the lens (see e.g.~\cite{Keeton:2001ss}).

We choose the characteristic lens scale $\xi_0$ to be the Einstein radius of the naked PBH, 
\be\label{eintein_radius}
    \xi_0 = \sqrt{\frac{4 m_{\rm PBH} D_l D_{ls}}{D_s}} \approx 0.014{\rm pc} \sqrt{\frac{m_{\rm PBH}}{M_\odot} \frac{D_s}{\rm Gpc} \frac{D_l D_{ls}}{D_s^2}},
\ee
which generally is smaller than the PDM minihalo radius~\eqref{radius_halo}. The deflection potential induced by the PBH dressed with the DM minihalo with a density profile of Eq.~\eqref{eq:profile} is $\psi(x) = \ln(x) + \psi_h(x)$, where the first term corresponds to the contribution from the PBH, which we describe as a point mass, and the second term is the contribution from the PDM minihalo around the PBH,
\bea
    \psi_h(x) \approx& \,12 (1-f_{\rm PBH}) \left[\frac{m_{\rm PBH}}{M_\odot}\right]^{\!\frac14} \left[\frac{R_h}{\rm pc}\right]^{\frac34} \\
    &\times \begin{cases} 
    \left[\frac{x \xi_0}{R_h}\right]^{\frac34} \,, \quad \quad \quad x \xi_0 \leq R_h \\
    1 + \frac34 \ln\!\left[\frac{x\xi_0}{R_h}\right] \,, \,\, x \xi_0 > R_h
   \end{cases} .
\eea
For the numerical computation of $F$, we use the method described in~\cite{Takahashi:2004phd} and, for $wy>5$, we switch to the geometric limit (see e.g.~\cite{Takahashi:2003ix}). 

\begin{figure}
\centering
\includegraphics[width=\columnwidth]{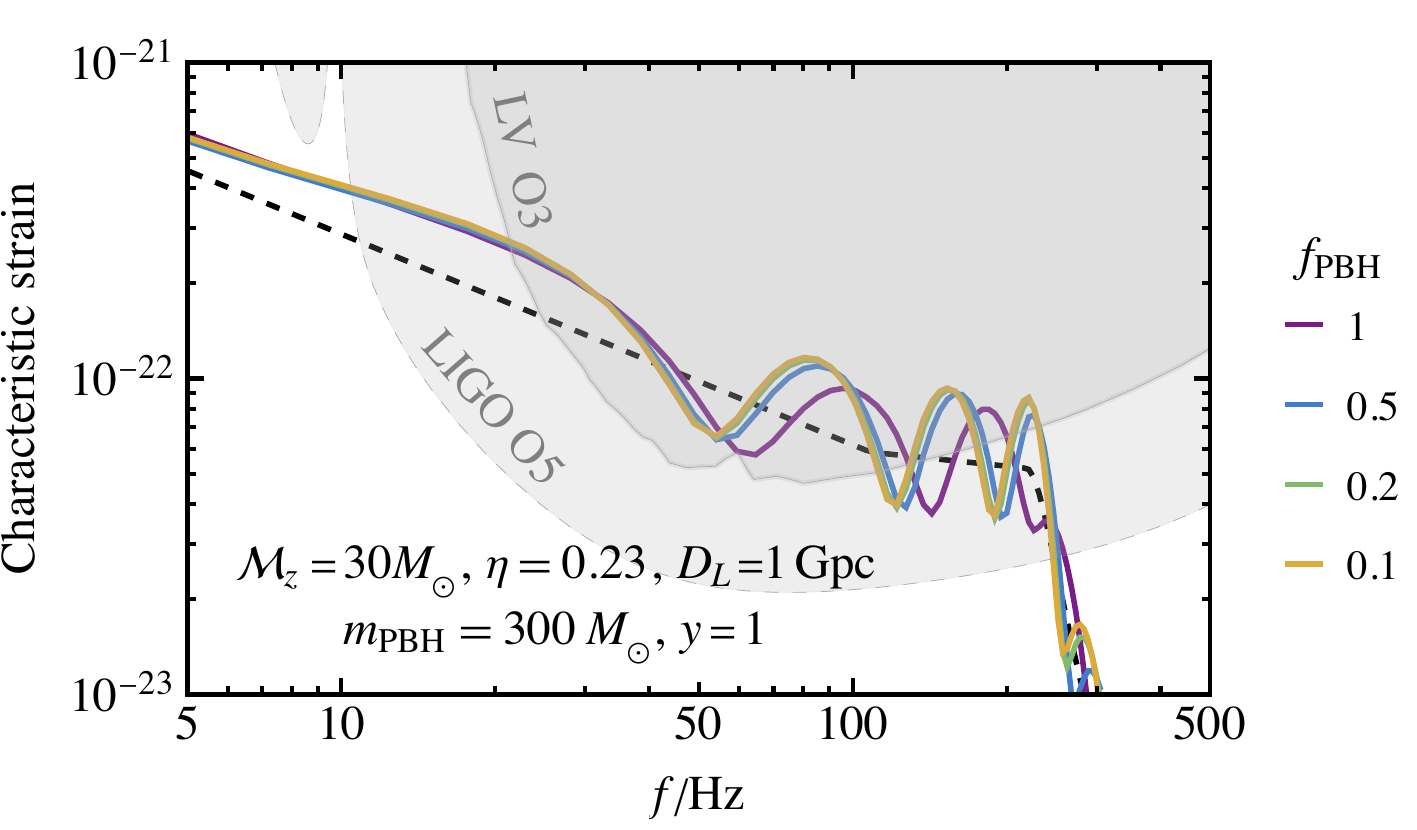}
\caption{The dimensionless characteristic GW strain, $\sqrt{f} |\tilde\phi(f)|$, from a BH binary merger. The black dashed curve shows the unlensed signal and the solid curves show the effect of a $300 M_\odot$ PBH lens  dressed in a PDM minihalo for different fractions of PBHs as DM. The shaded regions show the LIGO-Virgo O3~\cite{LIGOScientific:2021djp} and LIGO O5 (A+)~\cite{LIGOAplus} sensitivities.}
\label{fig:waveforms}
\end{figure}

In Fig.~\ref{fig:waveforms} we show examples of the lensed signal for various values of $f_{\rm PBH}$. The variations in the interference effect that modulates the amplitude of the signal are clearly visible when compared to the unlensed signal shown by the dashed curve. Moreover, we see that this effect depends on the fraction of DM in PBHs, as it, together with the PBH mass, determines the size of the PDM minihalo around the PBH.

\section{Analysis}
\label{sec:analysis}

\subsection{Detectability of the lens}

The Fourier transform of the GW signal can be expressed as $\tilde\phi(f) = A(f;\vect\theta) e^{-i\Psi(f;\vect\theta)}$. We compute the amplitude $A(f;\vect\theta)$ of the unlensed signal using the inspiral-merger-ringdown template~\cite{Ajith:2007kx}
\bea
    A&(f;\vect\theta) = \sqrt{\frac{5}{24}}\frac{\mathcal{M}_z^{\frac{5}{6}}}{\pi^{\frac{2}{3}}D_L}\\
    &\times \begin{cases} 
      f^{-\frac{7}{6}} & f<f_{\rm merg} \\
      f_{\rm merg}^{-\frac12}f^{-\frac23} & f_{\rm merg}\leq f< f_{\rm ring} \\
      f_{\rm merg}^{-\frac12}f_{\rm ring}^{-\frac23}\frac{\sigma^2}{4(f-f_{\rm ring})^2+\sigma^2} & f_{\rm ring}\leq f<f_{\rm cut} 
    \end{cases}
\eea
and the phase as
\be
\Psi(f;\vect\theta) = 2\pi f t_c - \phi_c + \frac{3}{128} (\pi \mathcal{M}_z f)^{-\frac53} \,.
\ee
Here $\mathcal{M}_z$ denotes the redshifted chirp mass of the binary, $\mathcal{M}_z \equiv (1+z) (m_1 m_2)^{\frac35}/(m_1+m_2)^{\frac15}$, and $D_L$ its luminosity distance. The frequencies $f_{\rm merg}$, $f_{\rm ring}$, $f_{\rm cut}$ and $\sigma$ are parameterised as $f_j = \eta^{\frac35} (a_j \eta^2 + b_j \eta + c_j)/(\pi \mathcal{M}_z)$, where $\eta$ denotes the symmetric mass ratio of the binary, $\eta \equiv m_1 m_2/(m_1+m_2)^2$, and $a_j$, $b_j$ and $c_j$ are coefficients whose fitted values are given in Table~I of~\cite{Ajith:2007kx}. This unlensed GW template includes five parameters: $\vect\theta = \{\mathcal{M}_z, \eta, D_L, t_c, \phi_c \}$. 

We determine the detectability of the lens by computing the log-likelihood difference (see e.g.~\cite{Fairbairn:2022xln})
\bea \label{eq:chi2}
    \Delta\chi^2(\vect\theta_s) = 4 \bar{\omega} \min_{\vect\theta} \int \!\td f \,\frac{|\tilde\phi_L(f;\vect\theta_s)-\tilde\phi_T(f;\vect\theta)|^2}{S_{\rm n}(f)} \,,
\eea
where $\tilde\phi_L(f;\vect\theta_s)$ is the lensed GW signal, $\tilde\phi_T(f;\vect\theta)$ the unlensed GW template, the minimum is taken over the parameters of the unlensed GW template, and $S_n(f)$ is the noise power spectral density, including the intrinsic detector noise and the noise from undetectable compact object binaries~\cite{Lewicki:2021kmu}. The prefactor $\bar{\omega}$ accounts for the sky location and the inclination of the binary, and the polarization of the signal. Averaging leads to $\bar\omega = 4/25$ and $\bar\omega = 3/25$ for L- and triangular-shaped detectors.

We compute $\Delta \chi^2(\vect\theta_s)$ numerically and find the maximal impact parameter $y_{\rm max}(\vect\theta_s)$ so that $\Delta \chi^2(\vect\theta_s) > 11.31$ for $y<y_{\rm max}(\vect\theta_s)$. This implies that for $y<y_{\rm max}(\vect\theta_s)$ the lensed template fits the signal better than the unlensed template at the $2\sigma$ confidence level. 

We can further use the log-likelihood test~\eqref{eq:chi2} to analyze whether different lens profiles can be experimentally resolved. To do this, we consider the lensed waveform characterized by the parameters $\vect\theta = \{\mathcal{M}_z, \eta, D_L, t_c, \phi_c, y, m_{\rm PBH},f_{\rm PBH} \}$ as the template $\tilde\phi_T(f;\vect\theta)$. For this 8-parameter fit, the $2\sigma$ confidence level corresponds to $\Delta\chi^2(\vect\theta_s) = 15.79$.

\subsection{Lensing probability}

The computation of $y_{\rm max}(\vect\theta_s)$ is of crucial importance since it determines the lensing cross section $\sigma(\vect\theta_s) = \pi \xi_0^2 y_{\rm max}(\vect\theta_s)^2$. The probability of a GW signal to be lensed follows from the probability of a lens object lying within the cross-section $\sigma$ from the line-of-sight to the GW source. It is given by $P_l(\vect\theta_s) = 1-e^{-\tau(\vect\theta_s)}$ where 
\be \label{eq:depth}
    \tau(\vect\theta_s) = \int_0^{z_s} \td z_l\, \frac{\sigma(\vect\theta_s) n(z_l)}{(1+z_l)H(z_l)}
\ee
is the optical depth. The number density of the lens objects is $n(z_l) = (1+z_l)^3 f_{\rm PBH} \rho_{\rm DM}/m_{\rm PBH}$.

We note that in~\eqref{eq:depth} we have assumed that PBHs were uniformly distributed in space. However, they would have inevitably clustered during the structure formation~\cite{Raidal:2018bbj, Hutsi:2019hlw, Inman:2019wvr, DeLuca:2020jug}. Studies of the effect of PBH clustering on optical microlensing~\cite{Petac:2022rio, Gorton:2022fyb} suggest the effect is negligible on the resulting constraints, except in cases where PBHs were initially formed in very compact clusters. The latter case would, however, clash with Lyman-$\alpha$ observations~\cite{DeLuca:2022uvz}. In short, our results remain robust also in the presence of observational allowed relatively mild PBH clustering. 

The expected number of lensed events from existing observations is given by
\be
    \bar{N}_l(m_{\rm PBH},f_{\rm PBH}) = \sum_{j=1}^{N} P_l(\vect{\theta}_j) \,,
\ee
where $N$ is the number of observed events. The prospects for future observations, on the other hand, can be obtained by integrating the expected number of lensed events from a model of the merger rate:
\be \label{eq:Nlens}
   \bar{N}_l(m_{\rm PBH},f_{\rm PBH}) = \mathcal{T} \!\int\! \td \lambda\, P_l(\vect\theta_s) \, p_{\rm det}\!\left(\frac{{\rm SNR}_c}{{\rm SNR}(\vect\theta_s)}\right) \,,
\ee
where $\mathcal{T}$ is the observation time, $p_{\rm det}$ is the detection probability~\cite{Gerosa:2019dbe}, SNR is the signal-to-noise ratio~\cite{Finn:1992wt} and
\be
    \td \lambda = \frac{1}{1+z_s}\frac{\td R}{\td m_1\td m_2} \frac{\td V_c}{\td z_s} \td m_1 \td m_2 \td z_s
\ee
is the differential rate of compact binary mergers. 
We use ${\rm SNR}_c = 8$ as the threshold SNR. The SNR is computed for an optimally oriented source-detector system and the averaging over the binary sky location and inclination and the polarization of the signal is accounted for by the detection probability $p_{\rm det}$.

We parametrize the merger rate of stellar mass BH binaries as
\be \label{eq:R}
    \frac{\td R}{\td m_1 \td m_2} = \frac{R_0}{Z_\psi} \eta^\beta \psi(m_1) \psi(m_2) {\rm SFR}(z) \,,
\ee
assuming redshift dependence proportional to the star formation rate~\cite{Belczynski:2016obo}.
At $z < 2$ this behaves roughly as $\propto(1+z)^{2.7}$, which is in good agreement with the LIGO-Virgo results~\cite{LIGOScientific:2021psn}. The factor $Z_\psi$ is chosen such that $R(z=0) = R_0$. We consider a cut double power-law shape for the BH mass function, 
\be
    \psi(m) \propto 
    \begin{cases}
        (m/m_2)^{\gamma_1}\, ,  & m_1 \leq m < m_2 \\
        (m/m_2)^{\gamma_2}\, ,  & m \geq m_2 \,,
    \end{cases}
\ee
and fix $m_1 = 3M_\odot$ and $m_2 = 55M_\odot$, corresponding to the lower and higher mass gaps. We fit the rest of the parameters, $R_0$, $\beta$, $\gamma_1$ and $\gamma_2$, to the LIGO-Virgo BH binary merger data~\cite{LIGOScientific:2018mvr,LIGOScientific:2020ibl,LIGOScientific:2021djp} following Ref.~\cite{Hutsi:2020sol}. In the following, we use the best fit values $R_0 = 15 \,{\rm yr}^{-1}{\rm Gpc}^{-3}$, $\beta = 4.8$, $\gamma_1 = -1.8$ and $\gamma_2 = -6.0$.

When assessing prospective sensitivities, we assume that the astrophysically motivated model~\eqref{eq:R} can describe the GW sources. However, if the abundance of stellar mass PBHs is sufficiently large, $f_{\rm PBH} \gtrsim 10^{-4}$, detectable PBH merger events are expected~\cite{Hutsi:2020sol, Franciolini:2021tla}. As we fix the magnitude of the merger rate from existing observations, a partially primordial origin of the GW sources would mainly imply that the merger rate grows monotonously with redshift, unlike the star formation rate. We will, however, not consider this possibility to ensure conservative constraints that are independent of the PBH merger rate. We stress also that the uncertainties related to the merger rate model do not apply to our analysis of the observed LIGO-Virgo events.

\section{Results and discussion}
\label{sec:results}

\begin{figure}
\centering
\includegraphics[width=0.95\columnwidth]{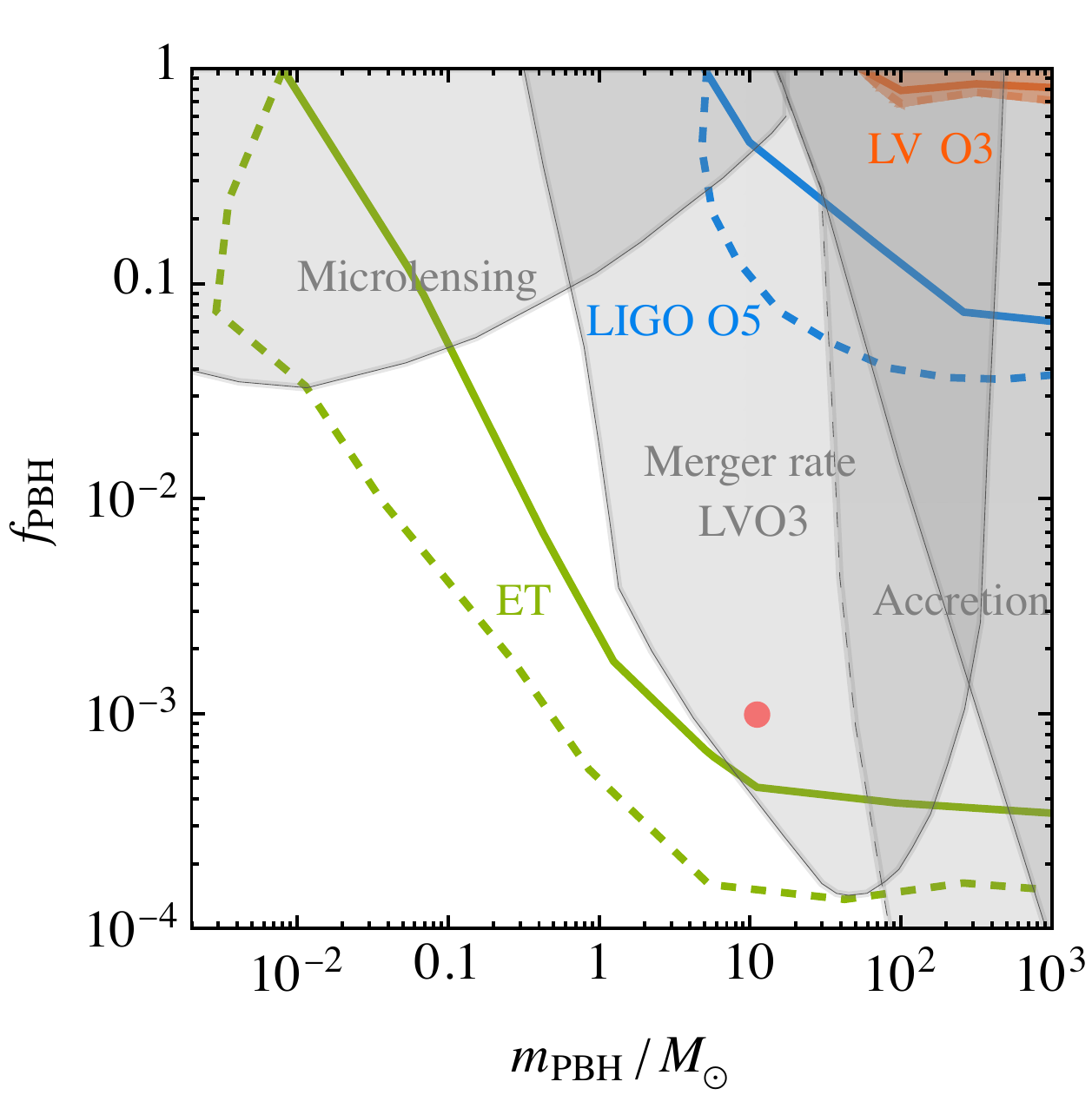}
\caption{The blue and green curves show the projected sensitivities of LIGO O5 and ET, and the orange curve shows the region disfavoured at $1\sigma$ by the current (O3) LIGO-Virgo observations. The gray curves show PBH constraints from microlensing~\cite{EROS-2:2006ryy}, merger rate measurements~\cite{Hutsi:2020sol}, and accretion~\cite{Serpico:2020ehh}. The solid and dashed curves show the results with and without the effect of the PDM minihalo. The red dot corresponds to the scenario in which all observed BH mergers would be primordial~\cite{Hutsi:2020sol, Hall:2020daa, Franciolini:2021tla}.}
\label{fig:prospects}
\end{figure}

During the first three observing runs, the LIGO-Virgo network detected a total of 90 compact binary coalescence events. We take source parameters and the corresponding sensitivity curves of the detectors from the LIGO-Virgo catalogs~\cite{LIGOScientific:2018mvr,LIGOScientific:2020ibl,LIGOScientific:2021djp}. For each event, we scale the sensitivity $S_n(f)$ so that the computed SNR matches the mean network SNR value reported in the catalogs. Then, using the scaled $S_n(f)$ and taking $\bar\omega = 1$ in Eq.~\eqref{eq:chi2}, we compute the lensing probability for each of these events as a function of $m_{\rm PBH}$ and $f_{\rm PBH}$, including the DM minihalos around the PBHs. This procedure accounts for the sky location, polarization and inclination of the binary, and uncertainties in our estimate of $S_n(f)$. When summing over the LIGO-Virgo events, we find that the expected total number of lensed events is $\bar{N}_l \approx 1.75$ if $m_{\rm PBH} > 100M_{\odot}$ and $f_{\rm PBH}=1$, and $\bar{N}_l \approx 0.5$ if $f_{\rm PBH}=0.2$ or if $f_{\rm PBH}=1$ and $m_{\rm PBH}\approx 20M_{\odot}$. The LIGO-Virgo collaboration has searched for lensing signatures in its catalog with negative results~\cite{LIGOScientific:2021izm}. So, the expected number of lensed events implies a mild $1\sigma$ tension with $f_{\rm PBH} \geq 0.7$ for $m_{\rm PBH}>100 M_{\odot}$ for dressed PBHs and $f_{\rm PBH} \geq 0.8$ for naked PBHs and $f_{\rm PBH} < 1$ for $m_{\rm PBH}>50 M_{\odot}$ in both scenarios, as shown by the orange curves in Fig.~\ref{fig:prospects}. Our constraint is weaker than estimated in Ref.~\cite{Urrutia:2021qak}, but can reach lower masses. This is because the analysis in~\cite{Urrutia:2021qak} was based on the accuracy at which the lensed waveform parameters can be measured, but it did not consider whether a lensed template gives a significantly better fit than an unlensed one.

We estimate the sensitivity of the LIGO O5 and ET to the PBH abundance by using the BH binary merger rate model~\eqref{eq:R}.\footnote{
As a cross-check, we derived the LIGO-Virgo O3 constraint using the merger rate~\eqref{eq:R} instead of the observed individual events and found a very good agreement between the results.} The prospects are shown by the green and blue curves in Fig.~\ref{fig:prospects}.\footnote{Fig.~\ref{fig:prospects} is derived assuming a monochromatic PBH mass function. The results for extended PBH mass functions can be computed using the method described in Ref.~\cite{Carr:2017jsz}.} We find that LIGO O5 can probe $f_{\rm PBH} \geq 0.1$ for masses $m_{\rm PBH}>10 M_{\odot}$, and ET can reach abundances $f_{\rm PBH} = 10^{-3}$ and subsolar mass PBHs with masses $m_{\rm PBH} \gtrsim 0.01 M_{\odot}$. Importantly, abundances of the order $f_{\rm PBH} = 10^{-3}$ correspond to the scenario in which a significant fraction of the observed BH-BH merger events were primordial~\cite{Hutsi:2020sol, Hall:2020daa, Franciolini:2021tla}. So, GW lensing with ET can provide a robust independent probe for PBH scenarios of the observed BH binary mergers. 

As shown in Fig.~\ref{fig:prospects}, the PDM dress enhances the sensitivity to $f_{\rm PBH}$ by a factor of $\mathcal{O}(1-10)$ and the effect is slightly stronger at higher $f_{\rm PBH}$.\footnote{We find that the PDM minihalo has a much weaker effect than was suggested in the recent strong lensing study~\cite{Oguri:2022fir}. Apart from differences in the simplifications, this discrepancy is mostly because a much denser PDM minihalo was used in~\cite{Oguri:2022fir}.} Notably, the enhancement in the sensitivity at $f_{\rm PBH} \leq 10^{-2}$ is smaller than one would naively expect from the PDM minihalo mass~\eqref{eq:Mh} because PDM minihalo is much larger than the typical Einstein radius for low $f_{\rm PBH}$. We remark that our constraints can also be extended to sufficiently compact extended objects, like DM stars, as was discussed in~\cite{Fairbairn:2022xln}.

\begin{figure}[h!]
\centering
\includegraphics[width=\columnwidth]{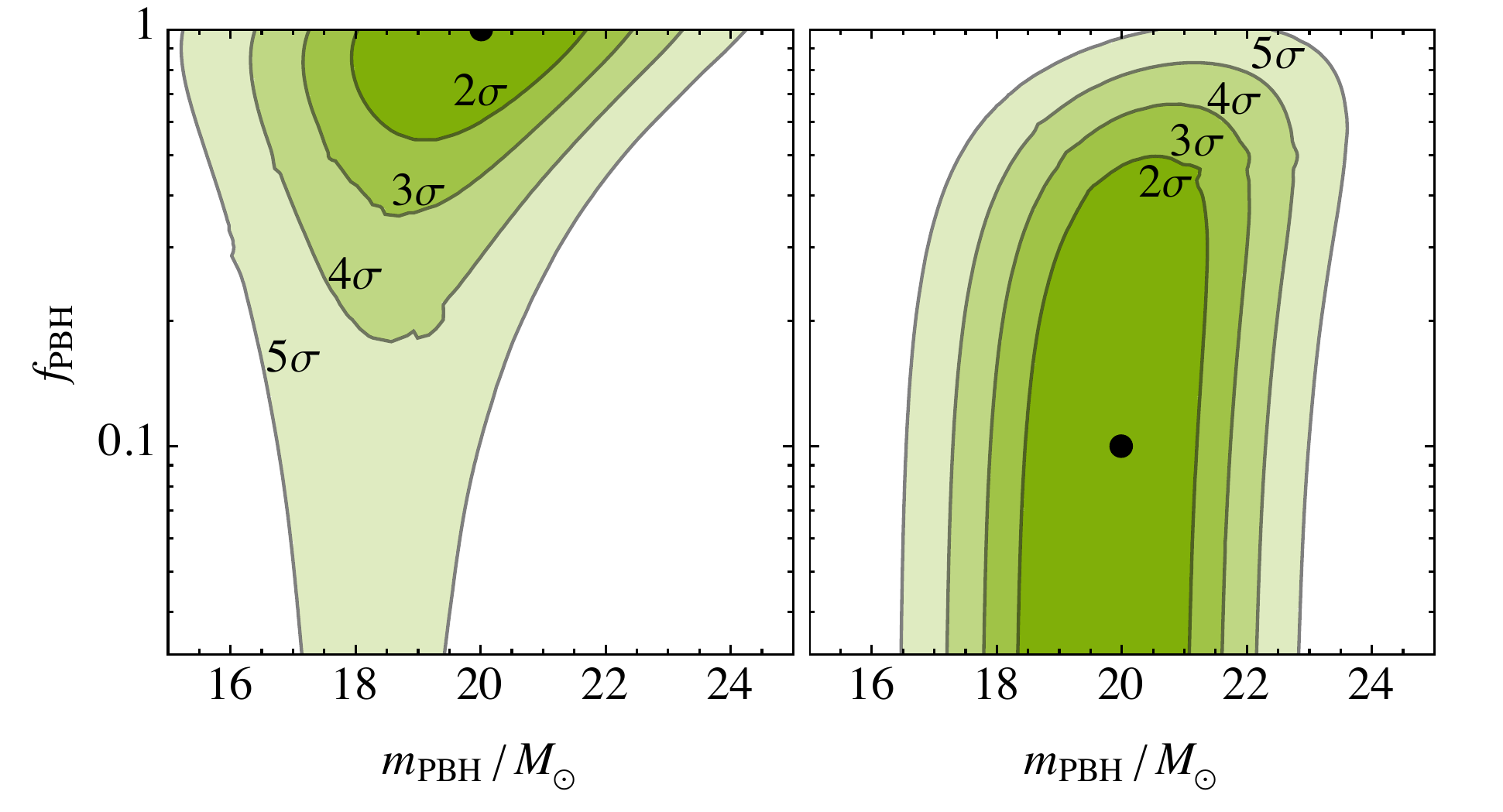}
\caption{Sensitivity of ET to the lens profile for two benchmark cases with $\mathcal{M}_z=40 M_\odot$, $\eta=0.23$, $D_L=5$\,Gpc, $m_{\rm PBH} = 20 M_\odot$ and $y=5$. The GW signal is assumed to correspond to a naked PBH lens ($f_{\rm PBH} = 1$) in the left panel and to a dressed lens (with $f_{\rm PBH} = 0.1$) in the right panel. The shading shows parameter regions in which ET can distinguish the assumed waveform (black dot) from a lensed signal characterized by $(m_{\rm PBH}, f_{\rm PBH})$ at the $2\sigma-5 \sigma$ confidence levels.}
\label{fig:pm_vs_halo}
\end{figure}

Finally, we investigate whether a heavier point mass can mimic a dressed PBH. In previous studies, the dressed PBH was modeled as a point mass lens whose mass was equal to the integrated mass up to the Einstein radius~\cite{Oguri:2022fir}. If this approximation fails, analyzing a lensed event could potentially reveal information about the environment surrounding the BH. We find that LIGO O5 is unlikely to see events for which dressed and naked lenses can be distinguished. On the other hand, ET is expected to see several sufficiently high SNR events allowing it to resolve the lens profile. Two realistic benchmark cases where ET would be able to measure the PDM dress of a PBH lens are shown in Fig.~\ref{fig:pm_vs_halo}, illustrating that the presence or absence of the dress may be excluded at the $4\sigma$ confidence level. Such events can shed light on the origin of BHs in the Universe and, if the lens is interpreted as a PBH, would give a probe of $f_{\rm PBH}$ through the measurement of the PDM minihalo properties. In this way, GW lensing can be a powerful tool in constraining potential PBH scenarios, especially when combined with other observables such as the PBH binary merger rate.

\section{Conclusions}

We have examined gravitational wave microlensing by primordial black holes, accounting for the effect of the particle dark matter minihalos when primordial black holes make up only a fraction of dark matter. We have further improved the existing analyses by including the full lens profile, employing a more complete waveform that contains the inspiral, merger, and ringdown phases, and performing a matched filtering analysis of the detectability of the lens.

We find that the existing LIGO-Virgo observations disfavor $f_{\rm PBH} > 0.7$ at $1\sigma$ in the mass range $m_{\rm PBH} \gtrsim 50M_{\odot}$. Moreover, we have shown that future observing runs of LIGO will be sensitive to $f_{\rm PBH} \gtrsim 0.1$ and $m_{\rm PBH} > 5 M_{\odot}$ and next-generation gravitational wave observatories like ET will be sensitive $f_{\rm PBH} \gtrsim 2 \times 10^{-4}$ and $m_{\rm PBH} \gtrsim 0.01 M_{\odot}$. In particular, they have the potential to rule out or confirm the primordial origin of the observed black hole merger events independently of the black hole binary merger rate measurements. We have also shown that ET can differentiate between dressed and naked black holes through gravitational wave lensing, providing an independent probe of the origins of black holes in the Universe and the particle dark matter properties, in case a lensed gravitational wave event would be detected.

Our results are expected to remain valid even if the standard minihalo profile and primordial black hole cosmology are modified in realistic ways (e.g. mild initial clustering). In summary, our study offers new insights into GW lensing by primordial black holes and opens up new avenues for exploring the phenomenology of dark matter and primordial black holes.

\begin{acknowledgments}
We thank Gert H\"utsi for useful discussions. This work was supported by European Regional Development Fund through the CoE program grant TK133 and by the Estonian Research Council grants PRG803 and PSG86. The work of V.V. has been partially supported by the European Union's Horizon Europe research and innovation program under the Marie Sk\l{}odowska-Curie grant agreement No. 101065736.
\end{acknowledgments}

\bibliography{refs}

\end{document}